\documentclass[prb,aps,twocolumn,showpacs,preprintnumbers,amsmath,amssymb]{revtex4}

\usepackage{graphicx}
\usepackage{dcolumn}
\usepackage{bm}
\usepackage{booktabs}
\usepackage{multirow}
\usepackage{amssymb}

\begin{document}


\title{Anomalous Hall conductivity and electronic structures of Si-substituted Mn$_{2}$CoAl epitaxial films}

\author{K. Arima,$^{1}$ F. Kuroda,$^{1}$ S. Yamada,$^{1,2}$ T. Fukushima,$^{2,3,4}$ T. Oguchi,$^{2,5,6}$\footnote{E-mail: oguchi@sanken.osaka-u.ac.jp} and K. Hamaya$^{1,2}$\footnote{E-mail: hamaya@ee.es.osaka-u.ac.jp}}
\affiliation{%
$^{1}$Graduate School of Engineering Science, Osaka University, Toyonaka 560-8531, Japan}%
\affiliation{%
$^{2}$Center for Spintronics Research Network, Osaka University, Toyonaka 560-8531, Japan}%
\affiliation{%
$^{3}$Institute for NanoScience Design, Osaka University, Toyonaka, Osaka 560-8531, Japan}
\affiliation{%
$^{4}$Institute for Datability Science, Osaka University, Suita, Osaka 565-0871, Japan}
\affiliation{%
$^{5}$Institute of Scientific and Industrial Research, Osaka University, Ibaraki, Osaka 567-0047, Japan}%
\affiliation{%
$^{6}$MI2I, National Institute for Materials Science, Tsukuba 305-0047, Japan}%


\date{\today}

\begin{abstract}
We study anomalous Hall conductivity ($\sigma$$_{\rm AHC}$) and electronic band structures of Si-substituted Mn$_{2}$CoAl (Mn$_{2}$CoAl$_{1-x}$Si$_{x}$).
First-principles calculations reveal that the electronic band structure is like a spin-gapless system even after substituting a quaternary element of Si for Al up to $x = $0.2 in Mn$_{2}$CoAl$_{1-x}$Si$_{x}$. 
This means that the Si substitution enables the Fermi level shift without largely changing the electronic structures in Mn$_{2}$CoAl. 
By using molecular beam epitaxy (MBE) techniques, Mn$_{2}$CoAl$_{1-x}$Si$_{x}$ epitaxial films can be grown, leading to the systematic control of $x$ (0 $\le$ $x$ $\le$ 0.3). 
In addition to the electrical conductivity, the values of $\sigma$$_{\rm AHC}$ for the Mn$_{2}$CoAl$_{1-x}$Si$_{x}$ films are similar to those in Mn$_{2}$CoAl films shown in previous reports. 
We note that a very small $\sigma$$_{\rm AHC}$ of $\sim$ 1.1 S/cm is obtained for $x =$ 0.225 and the sign of $\sigma$$_{\rm AHC}$ is changed from positive to negative at around $x =$ 0.25. 
We discuss the origin of the sign reversal of $\sigma$$_{\rm AHC}$ as a consequence of the Fermi level shift in MCA. 
Considering the presence of the structural disorder in the Mn$_{2}$CoAl$_{1-x}$Si$_{x}$ films, we can conclude that the small value and sign reversal of $\sigma$$_{\rm AHC}$ are not related to the characteristics of spin-gapless semiconductors. 

\end{abstract}

\maketitle
\section{INTRODUCTION}
When the conduction and valence band edges meet at the Fermi level and there is no gap for one spin channel while there is a finite gap in another spin channel, these materials are so called spin gapless semiconductors (SGSs).\cite{Wang_PRL,Wang_NPG,Wang_JMCC2016} 
In the field of spintronics, because the SGSs have intriguing physical properties such as electric-field induced magnetization changes, these materials can be utilized as ferromagnetic semiconductors.\cite{Ohno} 
Also, since not only the electrons but also the holes can become fully spin polarized,\cite{Wang_PRL,Wang_NPG,Wang_JMCC2016} 
one can utilize these materials as highly efficient spin injectors and highly spin polarized channels with tunable magnetic properties. 

In recent years, Ouardi {\it et al.} experimentally demonstrated that an inverse Heusler compound Mn$_{2}$CoAl (MCA) shows SGS characteristics such as the vanishing Seebeck coefficient and positive nonsaturating magnetoresistance with a linear change.\cite{Ouardi_PRL2013} 
Also, the expected carrier mobility of the MCA bulk has reached $\sim$ 70000 cm$^{2}$/V$\cdot$s, calculated from the results in Ref. 5. 
Since the observations of these characteristics, lots of theories and experiments on SGS characteristics of Heusler compounds for both bulk and films have been reported.\cite{Xu_EPL2013,Skaftouros_APL2013,Gao_APL2013,Galanakis_JAP2014,Jakobsson_PRB2015,Bainsla_APR2016,Jamer_APL2013,Jamer_JAP2014,Xu_APL2014,Sun_AIP2016,UedaMCA} 
For MCA films, however, there is no report on the vanishing Seebeck coefficient and/or on the positive nonsaturating magnetoresistance. 
Also, there is a large difference in the carrier concentration between the bulk in Ref. 5 ($\sim$10$^{17}$ cm$^{-3}$) and films (10$^{21}$ cm$^{-3}$ $\sim$10$^{22}$ cm$^{-3}$).\cite{Jamer_APL2013,Xu_APL2014,Sun_AIP2016,UedaMCA}
These imply that SGS characteristics have not been demonstrated in the MCA films. 

On the other hand, a very small anomalous Hall conductivity ($\sigma$$_{\rm AHC}$) has also been considered to be a characteristic of MCA.\cite{Ouardi_PRL2013}  
With respect to the bulk MCA,\cite{Ouardi_PRL2013} a small $\sigma$$_{\rm AHC}$ of 21.8 S/cm was observed although conventional ferromagnetic Heusler compounds showed relatively large $\sigma$$_{\rm AHC}$ of $\sim$ 1000 S/cm.\cite{Bomber_PRL2013,Vidal_APL2011}
According to Ouardi {\it et al.},\cite{Ouardi_PRL2013} a numerical calculation based on the Berry curvature showed a $\sigma$$_{\rm AHC}$ of 3 S/cm, which was regarded as a consequence of the antisymmetry of the Berry curvature for ${\bf k_{z}}$ vectors of opposite sign.
First-principles calculations also revealed an extremely small $\sigma$$_{\rm AHC}$ of 0.16 S/cm for ideal (stoichiometric) MCA.\cite{Kudrnovsky_PRB2013} 
Namely, such small $\sigma$$_{\rm AHC}$ values due to intrinsic anomalous Hall effect are peculiar properties of MCA. 

Recently, relatively small $\sigma$$_{\rm AHC}$ values were shown experimentally even for MCA films.\cite{Jamer_APL2013,Xu_APL2014,Sun_AIP2016,UedaMCA} 
Some of studies regarded such small $\sigma$$_{\rm AHC}$ values as an evidence for the realization of SGS-like MCA.\cite{Xu_APL2014,UedaMCA}  
However, the experimentally reported $\sigma$$_{\rm AHC}$ values are scattered, and the relationship between the electrical conductivity ($\sigma$$_{xx}$) and the carrier concentration in MCA films can not be understood simply as SGS characteristics like a bulk. 
So far, there has been no discussion about the correlation between $\sigma$$_{\rm AHC}$ and its electronic structures for MCA.

\begin{figure}[b]
\includegraphics[width=8cm]{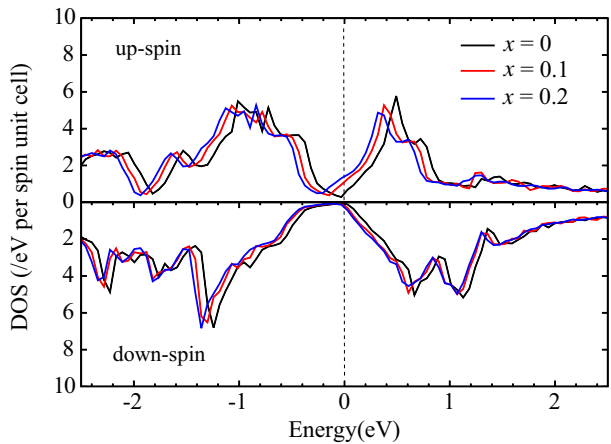}
\caption{(Color online) Spin-resolved DOS for the Si-substituted MCA, together with the perfectly ordered MCA. }
\end{figure}

In this article, to examine the correlation between $\sigma$$_{\rm AHC}$ and electronic band structures in MCA, we conduct first-principles calculations of Si-substituted Mn$_{2}$CoAl (Mn$_{2}$CoAl$_{1-x}$Si$_{x}$) and experimentally measure $\sigma$$_{\rm AHC}$ for the films with various $x$ (0 $\le$ $x$ $\le$ 0.3).
First-principles calculations reveal that the electronic band structure is like a spin-gapless (SG) system at around the Fermi level even after substituting a quaternary element of Si for Al up to $x =$ 0.2. 
This means that the Si substitution enables the Fermi level shift without largely changing the electronic structures in MCA. 
By using molecular beam epitaxy (MBE) techniques, Mn$_{2}$CoAl$_{1-x}$Si$_{x}$ (MCAS) epitaxial films can be grown. 
In addition to the electrical conductivity, the values of $\sigma$$_{\rm AHC}$ for the MCAS films are similar to those in MCA films shown in previous reports. 
We note that a very small $\sigma$$_{\rm AHC}$ of $\sim$ 1.1 S/cm is obtained for $x =$ 0.225 and the sign of $\sigma$$_{\rm AHC}$ is changed from positive to negative at around $x =$ 0.25. 
We discuss the origin of the sign reversal of $\sigma$$_{\rm AHC}$ as a consequence of the Fermi level shift in MCA. 
Because the carrier concentration of the grown MCAS films is five orders of magnitude larger than that of the bulk in Ref. 5, we can judge that the small value and sign reversal of $\sigma$$_{\rm AHC}$ are not related to the characteristics of spin-gapless semiconductors.

\section{Results}
\subsection{Electronic band structures}
To elucidate the electronic effect of Si substitution on the electronic structure of MCA, we use the MACHIKANEYAMA2002 program package,\cite{Akai} based on the Korringa-Kohn-Rostoker (KKR) Green's function method.\cite{Korringa,Kohn}  
The shape of the crystal potential is approximated with muffin-tin potentials and the angular momentum cutoff ($l_{\rm max}$) for the Green's function is $l_{\rm max} =$ 2. 
We employ generalized gradient approximation (GGA)\cite{Perdew} for the exchange-correlation functional. 
By total energy calculations, the equilibrium lattice constant of MCA is obtained to be $\sim$0.576 nm. 
Si substitution effects are considered within the coherent potential approximation (CPA) to treat the electronic structure and magnetism in disordered systems.\cite{Shiba,Soven} 
The Al sites of MCA are substituted with Si while keeping the lattice constant of pure MCA. 
In Fig. 1 we show the calculated spin-resolved density of states (DOS) for MCAS with $x =$ 0.1 and 0.2, together with the perfectly ordered MCA ($x =$ 0).
Even after substituting Si for Al up to $x =$ 0.2, the gapless structures in the majority-spin state can be maintained within a band gap in the minority-spin state, which are regarded as SG systems like pure MCA. 
When $x >$ 0.2 in MCAS, a SG system-like electronic structure was broken. 
Thus, the Si substitution up to $x =$ 0.2 can shift the position of the Fermi level for investigating $\sigma$$_{\rm AHC}$ in MCA systems. 

\subsection{Growth and characterizations}
MCAS films (0 $\le$ $x$ $\le$ 0.3) were grown on MgAl$_2$O$_4$(100) substrates by MBE, where the lattice mismatch between MCA (bulk) and MgAl$_2$O$_4$ was $\sim$1.5\%. 
After loading the MgAl$_2$O$_4$(100) substrates into an MBE chamber, we performed a heat treatment at 600 $^{\circ}$C for one hour with a base pressure of $\sim$10$^{-7}$ Pa. 
By $in-situ$ reflection high-energy electron diffraction (RHEED) observations, a good surface flatness of the MgAl$_2$O$_4$(100) substrate was confirmed.
Cooling the substrate temperature down to 300 $^{\circ}$C, we grew MCAS films with a thickness of $\sim$25 nm by co-evaporating Mn, Co, Al and Si using Knudsen cells. Here we used a nonstoichiometric evaporation technique,\cite{Hamaya_PRL, SYamada_APL,Fujita_PRAP} in which the evaporation ratio of Mn, Co, Al and Si is set to 2 : 0.68 : 2.2 : $x$ for MCAS. As an example, an {\it in-situ} RHEED pattern for $x$ = 0.2 is shown in the inset of Fig. 2(a), indicating good two-dimensional epitaxial growth of the MCAS film. 
\begin{figure*}[t]
\includegraphics[width=17cm]{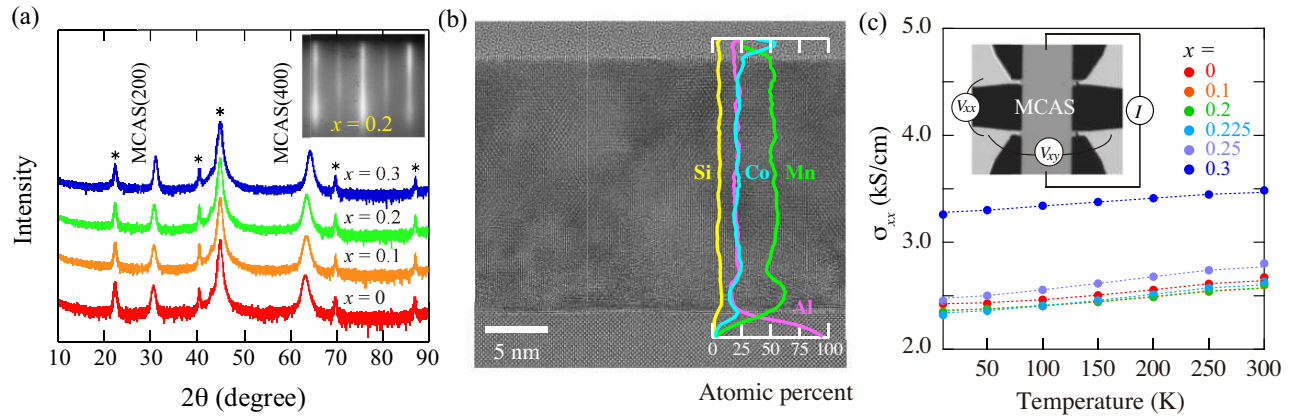}
\caption{(Color online) (a) $\theta$-2$\theta$ XRD patterns of the Mn$_2$CoAl$_{1-x}$Si$_{x}$ films. Asterisk marks denoted correspond to the diffraction peaks derived from the MgAl$_2$O$_4$(100) substrate. The inset shows a RHEED pattern of the surface for $x =$ 0.2 during the growth. (b) Cross-sectional TEM image of the Mn$_2$CoAl$_{0.8}$Si$_{0.2}$ film, together with the depth profile of the atomic compositions of Mn, Co, Al, and Si. (c) Temperature dependence of $\sigma$$_{xx}$ for various Mn$_2$CoAl$_{1-x}$Si$_{x}$ films. The inset shows a fabricated Hall-bar device for transport measurements. }
\end{figure*}

Figure 2(a) displays $\theta$-2$\theta$ x-ray diffraction (XRD) patterns of the MCAS films for various $x$. 
For all the films, (200) and (400) diffraction peaks are clearly observed at 2$\theta$ of $\sim$31$^{\circ}$ and 63$^{\circ}$, respectively, indicating the formation of $B2$-ordered Heusler alloys. 
No diffraction peak derived from the formation of other phases is observed. 
The estimated lattice constant of the grown MCA ($x =$ 0) was $\sim$0.587 nm, slightly larger than that of the bulk MCA ($\sim$0.579 nm)\cite{Ouardi_PRL2013} and the theoretical value ($\sim$0.576 nm) in our calculations. 
Because a half of diagonal length of MgAl$_2$O$_4$ is 0.5715 nm (1/$\sqrt{2}$ $\times$ 0.8083 nm), an in-plane lattice strain can be induced, resulting in an expansion of $c$-axis lattice constant for MCAS. 
In Fig. 2(a) we can see the slight shifts of the (200) and (400) diffraction peaks toward higher angles by increasing $x$, indicating that the $c$-axis lattice constant is gradually decreased with increasing $x$. 
This feature was consistent with the result obtained from CPA, indicating the successful substitution of Al for Si in the MCAS films. 

We further characterized an MCAS film by using cross-sectional transmission electron microscope (TEM) and energy dispersive x-ray spectroscopy (EDX).
From the TEM image in Fig. 2(b), we can recognize that the MCAS film for $x$ = 0.2 is epitaxially grown on MgAl$_2$O$_4$(100) with no marked defects and the surface of the MCAS film is very flat. 
The EDX line profiles of the MCAS/MgAl$_2$O$_4$(100) heterostructure are presented in the inset of Fig. 2(b). 
Even though we co-evaporated Mn, Co, Al and Si with a ratio of 2 : 0.68 : 2.2 : 0.2 for the growth of MCAS ($x$ = 0.2), the chemical composition of the MCAS layer along the vertical direction is nearly stoichiometric (Mn : Co : Al : Si = 2 : 1 : 0.8 : 0.2). These features are almost the same as our previous works.\cite{SYamada_APL,Fujita_PRAP}
For these reasons, the nonstoichiometric MBE technique used here enables us to systematically grow MCAS films on MgAl$_2$O$_4$(100) with a very flat surface. 

To measure electrical and magnetotransport properties, the MCAS films were patterned into Hall-bar devices with 80$\times$80 $\mu$m$^2$ in size, as shown in the inset of Fig. 2(c), by a conventional photolithography and Ar ion milling technique. Electrical conductivity ($\sigma$$_{xx}$) was measured by a standard dc four-terminal method. 
Figure 2(c) shows $\sigma$$_{xx}$ as a function of the external temperature for various $x$ (0 $\le$ $x$ $\le$ 0.3). 
A weak temperature dependence with a slight positive slope of $\sigma$$_{xx}$ (semiconducting type), almost equivalent to the bulk MCA reported previously,\cite{Ouardi_PRL2013} is observed for all the $x$. 
However, the value of $\sigma$$_{xx}$ for $x = 0.3$ is relatively large compared to those for other $x$. 
A possible reason of the difference in $\sigma$$_{xx}$ is the difference in the film quality. The detailed discussion is presented later. 
According to some theoretical calculations,\cite{Luo_JPD2008,Xin_Intermetallic2017} it has been revealed that $L2_1$-type (Mn-Co-Mn-Al) MCA is a ferromagnetic metal while {\it XA}-type (Mn$_{1}$-Mn$_{2}$-Co-Al) MCA shows a SGS.
From the results of Fig. 2(c), we can tentatively judge that the predominant structure of the grown MCA and MCAS films is the {\it XA}-type structure.
In particular, the $\sigma$$_{xx}$ value for $x$ = 0.2 and 0.225 at 10 K was $\sim$2340 S/cm, which is almost equivalent to that for the bulk MCA ($\sim$2250 S/cm at 10 K).\cite{Ouardi_PRL2013}
\begin{figure}[t]
\includegraphics[width=7cm]{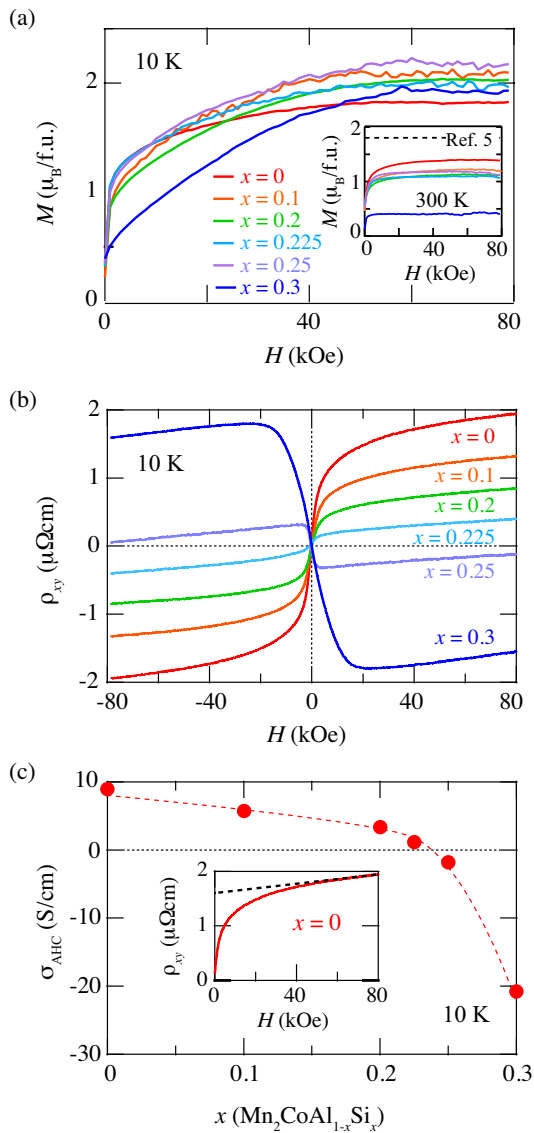}
\caption{(Color online) (a) Field-dependent magnetic moment ($M$) and (b) field-dependent Hall resistivity ($\rho$$_{xy}$) at 10 K for various MCAS films. The inset of (a) shows $M-H$ curves at 300 K. The dashed line is the value of $M$ for the bulk sample in Ref. 5. 
(c) $\sigma_{\rm AHC}$ as a function of $x$ at 10 K. The inset shows a method for the extrapolation of the value of $\rho$$_{xy}$ above 55 kOe towards zero field. }
\end{figure}

\subsection{Magnetic properties}
Magnetization curves of the various MCAS films (5 mm $\times$ 3 mm) were measured by applying in-plane magnetic fields ($H$). 
Here we used a vibrating sample magnetometer in a physical property measurement system (Quantum Design).
Figure 3(a) shows the plot of the magnetic moment ($M$) versus $H$ at 10 K. 
Because of the epitaxially grown films, there are some influences of the magnetocrystalline anisotropy on the magnetization curves in low-field regions. 
For all the films, the magnetic moments are saturated above 55 kOe, meaning that we can discuss the anomalous Hall effect above 55 kOe. 
The total magnetic moment at 10 K for all the films is nearly equivalent to that of bulk MCA reported previously.\cite{Ouardi_PRL2013} 
However, the theoretical magnetic moment, obtained from CPA, was slightly increased by increasing $x$ in MCAS. 
Also, the magnetic moments at 300 K for all the films are largely decreased compared to the bulk, as shown in the inset of Fig. 3(a). 
This means that the Curie temperature of the film samples is lower than that of bulk.\cite{ Xu_APL2014}
From these magnetic properties, we should consider the presence of the structural disorder in the MCAS films because of the low-temperature growth (300 $^{\circ}$C). 
Influence of the presence of the disorder is discussed in the last subsection. 


\subsection{Anomalous Hall effect}
Here we explore the anomalous Hall effect of the grown MCAS films. 
Figure 3(b) shows the field-dependent Hall resistivity ($\rho$$_{xy}$) at 10 K for various MCAS films (0 $\le$ $x$ $\le$ 0.3). 
All the films evidently show the anomalous Hall effect in addition to the ordinary Hall effect in high magnetic fields, and the feature is systematically changed with increasing $x$. 
To evaluate $\rho$$_{xy}$, we used a method for the extrapolation of the $\rho$$_{xy} - H$ curve above 55 kOe towards zero field, as shown in the inset of Fig. 3(c). 
From the relation, $\sigma_{yx}$ $\approx$ $\rho_{xy}$/$\rho_{xx}^{2}$,\cite{Nagaosa} we can estimate $\sigma$$_{\rm AHC}$ as the value of $\sigma_{yx}$ at zero field for various $x$, as shown in Fig. 3(c). 
For $x$ = 0 (MCA), a relatively small $\sigma$$_{\rm AHC}$ of $\sim$8.9 S/cm is obtained, comparable to that for thin-film samples reported elsewhere. \cite{Jamer_APL2013,Xu_APL2014,Sun_AIP2016,UedaMCA}
Thus, we can consider that the electronic band structure of the grown MCA film is similar to that expected from a numerical calculation of $\sigma$$_{\rm AHC}$ based on the Berry curvature \cite{Ouardi_PRL2013} or first-principles calculations.\cite{Kudrnovsky_PRB2013}
With increasing $x$, $\sigma$$_{\rm AHC}$ is gradually decreased and a very small $\sigma_{\rm AHC}$ of $\sim$1.1 S/cm is obtained for $x$ = 0.225. 
We note that a sign reversal of $\sigma$$_{\rm AHC}$ from positive to negative can be seen at around $x$ = 0.25. 
For $x$ = 0.3, the magnitude of negative $\sigma$$_{\rm AHC}$ is further increased. 
These systematic changes and sign reversal of $\sigma$$_{\rm AHC}$ have not been seen yet in MCA systems. 
\begin{figure}
\includegraphics[width=8.5cm]{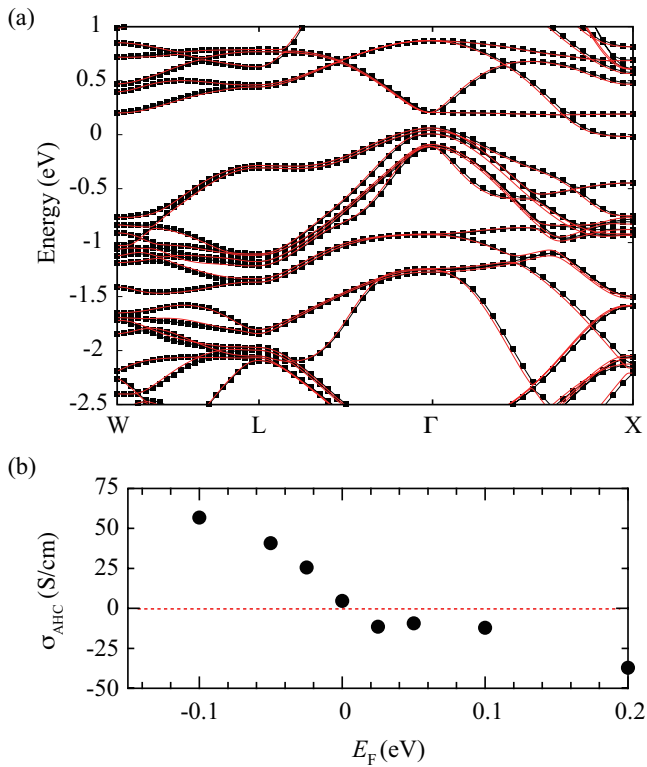}
\caption{(Color online) (a) Electronic band structure of MCA. Black dotted and red solid curves are the energy bands obtained from first-principles calculations and Wannier interpolation, respectively. (b) The plot of $\sigma$$_{\rm AHC}$, calculated from the Berry phase approach, versus the Fermi energy.}
\end{figure}

\section{Discussion}
\subsection{An origin of the AHC sign reversal}


From a theoretical point of view, we discuss an origin of the sign reversal of $\sigma$$_{\rm AHC}$, shown in Fig. 3, in the MCAS films. 
To calculate $\sigma$$_{\rm AHC}$, the QUANTUM ESPRESSO package\cite{Giannozzi} is used with the relativistic version of the ultrasoft pseudo-potentials using the GGA exchange-correlation functional.\cite{Perdew} 
The wave vector {\bf k} point mesh is taken to be 18 $\times$ 18 $\times$ 18, and Methfessel-Paxton smearing with a broadening parameter of 0.001 Ry is used. 
The cutoff energy for the wave function is set to 50 Ry. The Wannier interpolation models are obtained from the first-principles band structure using the Wannier90 program code.\cite{Marzari} 
For Mn and Co, $s$, $p$ and $d$ orbitals are adopted to construct the Wannier function. 
Figure 4(a) displays the calculated electronic band structure of the MCA with a lattice constant of 0.584 nm, where this value is one of the lattice constant values for a Si substituted MCA film ($x =$ 0.2) obtained experimentally. 
We can confirm that the electronic band structure is similar to that of MCA reported in previous works.\cite{Galanakis_JAP2014,Liu_PRB2008} 
Obtained Wannier interpolation energy band structure reproduces well the first-principles one as shown in the black dotted curve in Fig. 4(a). 
$\sigma$$_{\rm AHC}$ can be calculated using the Wannier interpolation models with following equation,\cite{Yao_PRL2004,Kubler_PRB2012,Wang_PRB} 
\begin{equation}
\sigma_{\rm AHC} = \frac{e^{2}}{\hbar}{\int}\frac{d^{3}{\bf k}}{(2\pi)^{3}}\Omega_{z}({\bf k}),
\end{equation}
where $\Omega_{z}$({\bf k}) is the $z$ component of the total Berry curvature for {\bf k}.
Figure 4(b) shows the calculated result of $\sigma$$_{\rm AHC}$ versus the Fermi energy ($E$$_{\rm F}$). 
Here the lattice constant is also assumed to be 0.584 nm, taken from our experimental result for $x =$ 0.2. 
At $E$$_{\rm F} =$ 0 (MCA), a $\sigma$$_{\rm AHC}$ of $\sim$ 4.6 S/cm is obtained theoretically. 
We note that the value of $\sigma$$_{\rm AHC}$ can be varied with shifting the position of $E$$_{\rm F}$, and the negative $\sigma$$_{\rm AHC}$ can be seen. 
In addition, from the results with virtual crystal approximation, the sign reversal of $\sigma$$_{\rm AHC}$ was confirmed in the Si substituted MCA.  
From these considerations, the Fermi level shift induced by the Si substitution in MCA enables the sign reversal of the $\sigma$$_{\rm AHC}$ value. 
Thus, the observed sign reversal of $\sigma$$_{\rm AHC}$, shown in Fig. 3, can roughly be interpreted in terms of the energy shift of the Fermi level in the Si-substituted MCA systems.

\subsection{Influence of Mn-Al antisite defects}

From the slope of the ordinary Hall effect above 55 kOe, the carrier (hole) concentration of our MCAS films can be estimated to be $\sim$10$^{22}$ cm$^{-3}$. 
This value is five orders of magnitude larger than that in bulk MCA ($\sim$10$^{17}$ cm$^{-3}$).\cite{Ouardi_PRL2013} 
Although the value and temperature dependence of $\sigma$$_{xx}$ for the MCAS films showed the semiconducting characteristics derived from the {\it XA}-type structure (Mn$_{1}$-Mn$_{2}$-Co-Al), we should consider the presence of the structural disorder causing a large number of additional carriers in the films. 
In a previous work on MCA,\cite{Galanakis_JAP2014} it was predicted that the SGS behavior cannot be conserved when there are antisite defects between Mn$_{2}$ and Al induced by the lattice strain and $etc$.
At least, since the grown MCAS films are structurally affected by the lattice strain from the MgAl$_2$O$_4$ substrate and by the low-temperature growth, we should check the influence of the antisite defects even in MCAS.  
In Fig. 5 we show the influence of the Mn$_{2}$-Al antisite defects on the spin-resolved DOS of MCAS with $x =$ 0, 0.1, and 0.2 using the KKR-CPA calculation, where we intentionally introduce 5 \% Mn$_{2}$-Al antisite defects. 
Unlike Fig. 1, the SGS-like electronic band structure is broken due to the formation of the majority spin band structure at around the Fermi level, giving rise to the half-metal-like one. 
We infer that this majority spin state can create a large number of carriers (holes), resulting in the unexpectedly large carrier concentration of $\sim$10$^{22}$ cm$^{-3}$. 
Because the expected sign reversal of $\sigma$$_{\rm AHC}$ in Fig. 4(b) was governed by the relatively large Fermi level shift in MCA, the influence of a small amount of the antisite defects on the sign reversal of $\sigma$$_{\rm AHC}$ might be small. 

We also measured magnetoresistance ($\rho_{xx} - H$) curve at 10 K. All the films showed negative magnetoresistance, being different from that reported in bulk MCA as mentioned in section I.\cite{Ouardi_PRL2013} 
The conventional negative magnetoresistance indicates that we have not obtained the SGS-like electronic band structure yet.\cite{Abrikosov_PRB}  
Furthermore, the reported magnetic properties in section II. C involved the presence of the disorder in MCAS films. 
For improving these issues, we should suppress such antisite defects and demonstrate realistic SGS electronic band structure even in the film samples. 

\begin{figure}[t]
\includegraphics[width=8cm]{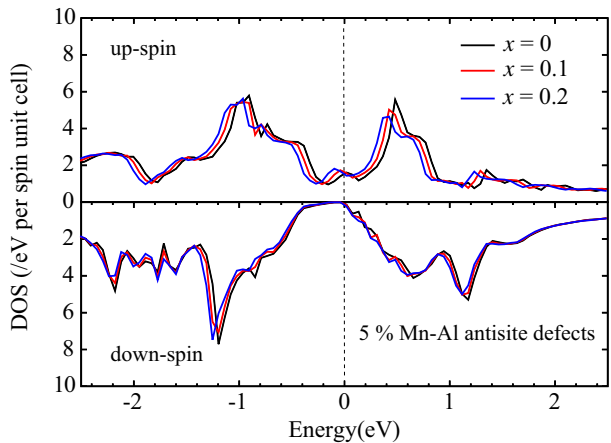}
\caption{(Color online) Spin-resolved DOS for MCAS ($x =$ 0, 0.1, and 0.2) with 5 \% Mn$_{2}$-Al antisite defects.}
\end{figure}

\section{Conclusion}
We have studied $\sigma$$_{\rm AHC}$ and electronic band structures of Si-substituted Mn$_{2}$CoAl (Mn$_{2}$CoAl$_{1-x}$Si$_{x}$).
First-principles calculations revealed that the electronic band structure is like a SG system even after substituting a quaternary element of Si for Al up to $x =$ 0.2. 
Using molecular beam epitaxy (MBE) techniques, we grew Mn$_{2}$CoAl$_{1-x}$Si$_{x}$ epitaxial films up to $x =$ 0.3. 
For $x \sim$ 0.225, a very small $\sigma$$_{\rm AHC}$ of $\sim$ 1.1 S/cm was obtained and the sign reversal of $\sigma$$_{\rm AHC}$ was seen at around $x =$ 0.25. 
We considered that one of the origins of the sign reversal of $\sigma$$_{\rm AHC}$ is the shift of the position of the Fermi level in the experimentally obtained MCA. 
However, the large number of carrier (hole) concentration of $\sim$10$^{22}$ cm$^{-3}$ could not be suppressed in the films, meaning that the SGS-like electronic band structure is broken. 
From now on, we should suppress the disorder such as antisite defects to achieve realistic SGS electronic band structure even in the film samples.

\section*{ACKNOWLEDGEMENTS}
This work was partly supported by a Grant-in-Aid for Scientific Research (A) (No. 16H02333) from the Japan Society for the Promotion of Science (JSPS). 


\end{document}